\documentclass [10pt]{article}
\usepackage{graphicx}

\begin{document}

\begin{center}
\textbf{Analytical forms of the deuteron wave function for Nijmegen group
potentials and polarization characteristics} \textbf{of
A(d,d')X reactions}
\end{center}

\begin{center}
\textbf{V. I. Zhaba}
\end{center}

\begin{center}
\textit{Uzhgorod National University, Department of Theoretical Physics,}
\end{center}

\begin{center}
\textit{54, Voloshyna St., Uzhgorod, UA-88000, Ukraine}
\end{center} 

\textbf{Abstract:} Polarization observables of the A(d,d')X
reactions have been calculated according to the received
coefficients of the analytical form for deuteron wave function in
coordinate space for the nucleon-nucleon Nijmegen group potentials
(NijmI, NijmII, Nijm93). The obtained values of tensor $A_{yy}$
and vector $A_{y}$ analyzing powers have been compared with the
published experimental data at $t$- scaling for the inelastic
scattering of deuterons on hydrogen, carbon and beryllium.
Theoretical values of tensor-tensor $K_{yy}$ and vector-vector
$K_{y}$ polarization transfers have been also evaluated in the
plane-wave impulse approximation.



\begin{center}
\textbf{1. Introduction}
\end{center}

Deuteron - is the simplest core, which consists of the two elementary
particles - a proton and a neutron. The simplicity and clarity of the
deuteron structure serves as a convenient laboratory for simulation and
analyzing nucleon-nucleon forces. Despite a detailed theoretical and
experimental study, the deuteron is of considerable interest today due to
certain theoretical inconsistencies. In particular, according to the review
\cite{ZhabaArxiv} in a number of papers, the deuteron wave function (DWF) in
coordinate representation has knots near the origin of coordinates. The
existence of such knots of the ground and the only state of the deuteron
attests to inconsistencies and inaccuracies in the implementation of
numerical algorithms in solving such problems, or to the features of
potential deuteron models \cite{Zhaba20161}.

In the review \cite{ZhabaArxiv}, the static parameters of the deuteron obtained
from DWF for various potential models have been systematized, and a review
of the analytical forms of DWF in coordinate representation has been carried
out. What is more, both analytical forms and parameterizations of DWF have
been presented, which are necessary for further calculations of the
processes characteristics with the participation of deuteron. In addition,
it has been noted that in such a convenient form, DWFs are needed when
applied to calculate the polarization characteristics of the deuteron, as
well as to evaluate the theoretical values of spin observables in
dp-scattering. After all, the DWF can also be used to calculate the
parameters and characteristics of the reactions of (d,d') type.

In this paper, analytical forms of DWF have been used for theoretical
calculations of a set of polarization observables in A(d,d')X reactions.
Realistic phenomenological Nijmegen group potentials (NijmI, NijmII and
Nijm93) \cite{Stoks1994, Swart1995} have been used for numerical calculations.

\begin{center}
\textbf{2. DWF analytical forms}
\end{center}

Among the large list of analytical forms of DWF in the coordinate
representation, parametrization of DWF for the Parisian potential
\cite{Lacombe1981} is worth highlighting

\begin{equation}
\label{eq1}
\left\{ {\begin{array}{l}
 u\left( r \right) = \sum\limits_{j = 1}^N {C_j \exp \left( { - m_j r}
\right),} \\
 w\left( r \right) = \sum\limits_{j = 1}^N {D_j \exp \left( { - m_j r}
\right)\left[ {1 + \frac{3}{m_j r} + \frac{3}{\left( {m_j r} \right)^2}}
\right],} \\
 \end{array}} \right.
\end{equation}

where $m_j = \beta + (j - 1)m_0 $; $\beta = \sqrt {ME_d } $,
$m_{0}$=0.9 fm$^{-1}$; $M$ - nucleon mass; $E_{d}$ - binding
energy of the deuteron. Extreme conditions at $r \to 0$:

\[
u\left( r \right) \to r;
\quad
w\left( r \right) \to r^3.
\]

The asymptotics of the deuteron wave function (\ref{eq1}) for $r \to \infty $

\[
\begin{array}{l}
 u(r) \sim A_S \exp ( - \beta r), \\
 w(r) \sim A_D \exp ( - \beta r)\left[ {1 + \frac{3}{\beta r} +
\frac{3}{(\beta r)^2}} \right], \\
 \end{array}
\]

where $A_{S}$ and $A_{D}$ are the asymptotics of $S$- and $D$- state normalizations.

The coefficients of the analytical form (\ref{eq3}) for $N>10$ are
determined by the formulas \cite{Lacombe1981}

\[
\left\{ {\begin{array}{l}
 C_n = - \sum\limits_{j = 1}^{n - 1} {C_j } \\
 D_{n - 2} = \frac{m_{n - 2}^2 }{\left( {m_n^2 - m_{n - 2}^2 } \right)\left(
{m_{n - 1}^2 - m_{n - 2}^2 } \right)}\left[ { - m_{n - 1}^2 m_n^2
\sum\limits_{j = 1}^{n - 3} {\frac{D_j }{m_j^2 } + \left( {m_{n - 1}^2 +
m_n^2 } \right)\sum\limits_{j = 1}^{n - 3} {D_j - \sum\limits_{j = 1}^{n -
3} {D_j m_j^2 } } } } \right] \\
 \end{array}} \right.
\]

or

\[
\sum\limits_{j = 1}^{N_b } {C_j } = 0;
\quad
\sum\limits_{j = 1}^{N_b } {D_j } = \sum\limits_{j = 1}^{N_b } {D_j m_j^2 }
= \sum\limits_{j = 1}^{N_b } {\frac{D_j }{m_j^2 }} = 0.
\]

The search for the coefficients of the analytical form (\ref{eq1}) was done for the
Paris \cite{Lacombe1981} and the Bonn (OBEPC \cite{Machleidt1989} and
charge-dependent Bonn (CD-Bonn) \cite{Machleidt2001}) potentials and the fss2
model (with the Coulomb exchange kernel \cite{Fujiwara2001}, and calculated
according to three different schemes (isospin basis and particle basis with
or without the Coulomb force and fss2 baryon-baryon interaction
\cite{Fukukawa2015}), with $N$=13, 11 and 11 respectively. What is more, formula (\ref{eq1})
was applied to MT model \cite{Krutov2007}, where $N_{S}$=16; $N_{D}$=12.

The new analytical DWFs were also proposed in coordinate representation in
the 2000s. These include, for instance, Dubovichenko's \cite{Dubovichenko2000}
and Berezhnoy-Korda-Gakh's \cite{Berezhnoy2005}, parameterizations, as well as
the analytic form in such a simple form \cite{Zhaba20162}

\begin{equation}
\label{eq2}
\left\{ {\begin{array}{l}
 u(r) = r\sum\limits_{i = 1}^N {A_i \exp ( - a_i r^2),} \\
 w(r) = r\sum\limits_{i = 1}^N {B_i \exp ( - b_i r^2).} \\
 \end{array}} \right.
\end{equation}

The DWF (\ref{eq1}) was used to approximate the numerical arrays
of radial wave functions obtained for Nijmegen group potentials
(NijmI, NijmII, Nijm93 and Reid93). The behavior of the $\chi ^{2
}$ value depending on the number of terms of the $N $expansion has
been studied. In addition to the DWF (\ref{eq2}), the analytical
form for Nijmegen group potentials (NijmI, NijmII and Nijm93) was
suggested in paper \cite{Zhaba20163}

\begin{equation}
\label{eq3}
\left\{ {\begin{array}{l}
 u(r) = r^{3 / 2}\sum\limits_{i = 1}^N {A_i \exp ( - a_i r^3),} \\
 w(r) = r\sum\limits_{i = 1}^N {B_i \exp ( - b_i r^3).} \\
 \end{array}} \right.
\end{equation}

Formula (\ref{eq3}) was used in paper \cite{Zhaba20161} for approximation of DWFs for the
Reid93 and Argonne v18 potentials. The obtained wave functions for these
potentials do not contain any superfluous knots. Theoretical evaluation of
the polarization characteristics of the deuteron using DWF (\ref{eq3}) is
appropriate.

DWF parameters in the dressed dibaryon model (DDM) \cite{Platonova20101} were
obtained for the analytical form \cite{Krasnopolsky1985}

\begin{equation}
\label{eq4}
\left\{ {\begin{array}{l}
 u(r) = r\sum\limits_{i = 1}^N {A_i \exp ( - a_i r^2),} \\
 w(r) = r^3\sum\limits_{i = 1}^N {B_i \exp ( - b_i r^2).} \\
 \end{array}} \right.
\end{equation}

\begin{center}
\textbf{3. Polarization observables of the A(d,d')X reactions}
\end{center}

Experimental determination of the values of polarization characteristics for
the deuteron fragmentation reaction of A(d,p)X and for (d,d') reactions of
inelastic scattering of the deuteron on nuclei \cite{Ladygin2002} remain one of
the main tools for studying the deuteron structure. In addition, the
reactions of inelastic scattering of deuterons are also used to study the
properties and the formation of baryonic resonances.

The singlet breakup contribution to the amplitude and others
elastic scattering observables for $^{58}$Ni(d,d)$^{58}$Ni
reaction at deuteron energy 400 MeV is studied in paper
\cite{AlKhalili1990}. The inclusion of the singlet channel
coupling produce a large effect on the calculated angular
distributions for reaction tensor analyzing power $A_{yy}$ but
insignificant effect on the cross section and vector analyzing
power $A_{y}$. It takes place through dynamically induced $T_{L}$
tensor interaction. Tensor and vector analyzing powers for
deuterons occurring from d+$^{58}$Ni inelastic scattering as a
function of outgoing deuteron energy are provided in
\cite{Stephenson1981}.

Within the framework of the plane-wave impulse approximation model (PWIA)
\cite{Ladygin2002} previously obtained experimental data \cite{Azhgirey1999} of tensor
and vector analyzing powers have been analyzed in the reaction of inelastic
scattering of deuterons on carbon at an initial momentum of deuteron of
9~GeV/c and a detection angle of secondary deuterons of 85~mrad in the
region of resonance excitation with a mass of 2190~MeV/c$^{2}$.

The results of experimental studies of the tensor $A_{yy}$ and vector
$A_{y}$ analyzing powers for inelastic scattering of deuterons with a
momentum of 4.5~GeV/c on beryllium at an angle of 80~mrad around the
excitation of baryonic resonances are provided in \cite{Ladygin2000}. The
$A_{yy}$ values are in good agreement with the experimental data at a zero
angle and show an approximate $t$- scaling up to $\sim $-0.9~(GeV/c)$^{2}$.
Paper \cite{Azhgirey2005} provides the experimental data of analyzing powers in
inelastic scattering of deuterons with a momentum of 5.0 GeV/c on beryllium
at an angle of 178 mrad in the vicinity of excitation of baryonic resonances
with a mass up to $\sim $1.8~GeV/c$^{2}$. Moreover, the values of $A_{yy}$
correlate well with the previous results at 4.5 and 5.5~GeV/c. The results
of these experiments are compared with the predictions of models with
multiple-scattering, PWIA and $\omega $-meson exchange models.

The angular dependence of analyzing powers in the inelastic
scattering of deuterons with a momentum of 9.0 GeV/c on hydrogen
and carbon has been measured in paper \cite{Ladygin2006}. The
measurement range corresponds to a baryonic resonance excitation
with a mass of 2.2-2.6~GeV/c$^{2}$. These data show approximate
$t$- scaling up to -1.5~(GeV/c)$^{2}$. The large figures for
$A_{y}$ value suggested a significant role of the spin-dependent
part of the elementary amplitude of the NN$ \to $NN* reaction.

The values of the tensor $A_{yy}$ and vector $A_{y}$ analyzing
powers can be received experimentally from the numbers of
deuterons $n^{+}$, $n^{-}$, $n^{0}$, registered for different
modes of beam polarization $p_{z}$, $p_{zz}$ and normalized to the
beam intensity, taking into account the dead-time effect of the
installation \cite{Azhgirey2005}:

\begin{equation}
\label{eq5}
A_{yy} = 2\frac{p_z^ - (n^ + / n^0 - 1) - p_z^ + (n^ - / n^0 - 1)}{p_z^ -
p_{zz}^ + - p_z^ + p_{zz}^ - }.
\end{equation}

\begin{equation}
\label{eq6}
A_y = - \frac{2}{3}\frac{p_{zz}^ - (n^ + / n^0 - 1) - p_{zz}^ + (n^ - / n^0
- 1)}{p_z^ - p_{zz}^ + - p_z^ + p_{zz}^ - }.
\end{equation}

Tensor and vector analyzing powers in PWIA are theoretically calculated
according to the following formulas \cite{Ladygin2002}:

\begin{equation}
\label{eq7}
A_{yy} = \frac{T_{00}^2 - T_{11}^2 + 4P^2T_{10}^2 }{T_{00}^2 + 2T_{11}^2 +
4P^2T_{10}^2 };
\end{equation}

\begin{equation}
\label{eq8} A_y = 2\sqrt 2 P\frac{(T_{11} +
T_{00})T_{10}}{T_{00}^2 + 2T_{11}^2 + 4P^2T_{10}^2 };
\end{equation}

where $T_{ij}(p/2)$ are the amplitudes determined by the radial wave functions of
the deuteron in coordinate representation $u(r)$ and $w(r)$:

\[
T_{00} = S_0 (p / 2) + \sqrt 2 S_2 (p / 2),
\]

\[
T_{11} = S_0 (p / 2) - \frac{1}{\sqrt 2 }S_2 (p / 2),
\]

\[
T_{10} = \frac{i}{\sqrt 2 }\int\limits_0^\infty {\left( {u^2 -
\frac{w^2}{2}} \right)j_0 dr} + \frac{i}{2}\int\limits_0^\infty {w\left( {u
+ \frac{w}{\sqrt 2 }} \right)j_2 dr} .
\]

Here $S_{0}$ and $S_{2}$ - spherical (charge) and quadrupole deuteron form
factors

\begin{equation}
\label{eq9}
S_0 (p / 2) = S_0^{(\ref{eq1})} + S_0^{(\ref{eq2})} ;
\end{equation}

\begin{equation}
\label{eq10}
S_2 (p / 2) = 2S_2^{(\ref{eq1})} - \frac{1}{\sqrt 2 }S_2^{(\ref{eq2})} ,
\end{equation}

which are expressed through elementary spherical $S_0^{(i)} $ and quadrupole
$S_2^{(i)} $ form factors \cite{Ladygin2002, Platonova20102} in the form

\begin{equation}
\label{eq11}
S_0^{(\ref{eq1})} = \int\limits_0^\infty {u^2j_0 dr} ;
\quad
S_0^{(\ref{eq2})} = \int\limits_0^\infty {w^2j_0 dr} ;
\end{equation}

\begin{equation}
\label{eq12}
S_2^{(\ref{eq1})} = \int\limits_0^\infty {uwj_2 dr} ;
\quad
S_2^{(\ref{eq2})} = \int\limits_0^\infty {w^2j_2 dr} .
\end{equation}

Here $j_{0}$, $j_{2}$ - spherical Bessel functions of zero and second order
from the \textit{pr}/2 argument; $p$ - momentum. Besides, formulas (\ref{eq11}) and (\ref{eq12}) have been
written down in paper \cite{Platonova20102} with \textit{pr} argument.

According to \cite{Ladygin2002}, a parameter $P = a \cdot p$ is introduced in the
formulas (\ref{eq7}) and (\ref{eq8}), which characterizes the ratio of the spin-dependent
and spin-independent parts of amplitudes for the NN$ \to $NN*(2190) process:

\begin{equation}
\label{eq13}
P = \frac{f^{sf}(p)}{f^{nf}(p)}.
\end{equation}

Here amplitudes $f^{sf}(p)$ and $f^{nf}(p)$ can be parametrized in such form
\cite{Ladygin2002}

\begin{equation}
\label{eq14}
f^{sf}(p) = i\sqrt {\frac{A_s }{\pi }} \exp \left( { - \frac{B_s p^2}{2}}
\right);
\end{equation}

\begin{equation}
\label{eq15}
f^{nf}(p) = i\sqrt {\frac{A_n }{\pi }} \exp \left( { - \frac{B_n p^2}{2}}
\right);
\end{equation}

where $A_{s}$, $B_{s}$, $A_{n}$, $B_{n}$ are constants. Values of constants
$A_{n}$ and $B_{n}$ can be determined from experimental data by excitation of
a resonance N*(2190) in pp- interactions. If parameters $B_{s}$ and $B_{n}$
for exponents coincide, then parameter $P$ will be linear function.

Within the $\omega $-meson exchange model \cite{Rekalo1996,
TomasiGustafsson1999}, the tensor analyzing power in deuteron inelastic
scattering is defined as \cite{Ladygin2000, Azhgirey2001}

\begin{equation}
\label{eq16}
A_{yy} = \frac{V_1^2 + (2V_0 V_2 + V_2^2 )\rho }{4V_1^2 + (3V_0^2 + V_2^2 +
2V_0 V_2 )\rho },
\end{equation}

where $\rho = \sigma _L / \sigma _T $ - ratio of the cross sections of
absorption of virtual isoscalar photons with longitudinal and transversal
polarizations by nucleons \cite{TomasiGustafsson1999}; the structure functions
$V_{0}$, $V_{1}$ and $V_{2}$ is expressed in terms of the electric $G_{C}$,
magnetic $G_{M}$ and quadrupole $G_{Q}$ deuteron form factors.

For N* resonance excitation, the ratio $\rho $ can be written in the form
\cite{TomasiGustafsson1999, Ladygin2000}

\begin{equation}
\label{eq17}
\rho _{N\ast } = \frac{\left| {A_l^p + A_l^n } \right|^2}{\left| {A_{1 /
2}^p + A_{1 / 2}^n } \right|^2 + \left| {A_{3 / 2}^p + A_{3 / 2}^n }
\right|^2},
\end{equation}

where $A_l^N $ - longitudinal form factor of the N* excitation on
a proton ($N=p)$ or a neutron ($N=n)$; $A_{1 / 2}^N $ ³ $A_{3 /
2}^N $ - two transversal form factors, corresponding to total
$\gamma $*+N helicity equal to 1/2 and 3/2. It has been specified
in paper \cite{Azhgirey2001} that the tensor analyzing power
$A_{yy}$ in $\omega $-meson exchange model in the $t$ channel can
be represented as the product of two parts that are determined by
the electromagnetic deuteron properties and by the form factors
for the N$ \to $N* transition.

Using the DWF (\ref{eq3}) for Nijmegen group potentials (the
coefficients of expansions are taken from paper
\cite{Zhaba20163}), the theoretical values of the tensor $A_{yy}$
and vector $A_{y}$ analyzing powers in PWIA have been calculated
by formulas (\ref{eq7}) and (\ref{eq8}), respectively. Moreover,
for the parameter $P$, the argument $a$=0.4 for $A_{yy}$ and
$a$=0.4$\div $1.1 for $A_{y}$ has been selected. The results of
numerical calculations are shown in Figures 1-4, where $t$-
scaling according to the elementary record is used according to
the formula $t$=-(0.197326$p)^{2}$ given units for momentum $p$ in
[fm$^{-1}$] and $t$- scaling in [(GeV/c)$^{2}$]. The theoretical
estimates of the $A_{yy}$ and $A_{y}$ values are compared with the
experimental data for the reaction of (d,d') type on light nuclei:
on hydrogen \cite{Ladygin2006}, carbon \cite{Ladygin2006,
Afanasiev1998, Azhgirey1998}, beryllium \cite{Ladygin2000,
Azhgirey2005, Azhgirey2001}. Based on the data
\cite{Azhgirey1998}, the tensor analyzing power is determined by
the formula: $A_{yy} = - T_{20} / \sqrt 2 $. In contrast to the
calculated value of the tensor analyzing power$ A_{yy}$, its
vector component $A_{y}$ strongly depends on the parameter $P$ and
better coincides with the experiment at $a$=0.4.

The tensor analyzing power $A_{yy}$ according to formula (\ref{eq7}) is weakly
dependent on the parameter $P$. If the spin-dependent part of the amplitude in
the formula for the parameter (\ref{eq13}) is equal to zero, then formula (\ref{eq7}) will
be written as \cite{Ladygin2002}

\begin{equation}
\label{eq18}
A_{yy} = \frac{1}{2}\frac{S_2^2 (p / 2) + 2\sqrt 2 S_0 (p / 2)S_2 (p /
2)}{S_0^2 (p / 2) + S_2^2 (p / 2)}.
\end{equation}

That is, tensor analyzing power $A_{yy}$ is determined only by the spherical
and quadrupole deuteron form factors. Calculations by formulas (\ref{eq7}) and (\ref{eq18})
practically coincide.

Papers \cite{Ladygin2002, Ladygin2006} provide the results of
calculations of the analyzing powers obtained by DWF for the Paris
and three (A, B, C) versions of the Bonn potentials. There, for
$a$=0.4, the value $A_{yy}$ intersects zero in the region $\vert
$\textit{t$\vert $}=1.0-1.2~(GeV/c)$^{2}$, and the vector capacity
$A_{y}$ calculated for the Paris potential better correlates with
the experimental data than for the Bonn potential ($a$=0.3-0.4),
and to $\vert $\textit{t$\vert $}=1.0~(GeV/c)$^{2}$ is determined
by the spin-dependent part of the amplitude of the elementary
process NN$ \to $NN*(2190).

It should be noted that the results of calculations of the
analyzing powers of $A_{yy}$ and $A_{y}$ as in papers
\cite{Ladygin2002, Ladygin2006} for the Paris and Bonn potentials,
and in this paper for Nijmegen group potentials, strongly differ
from the experimental data within $\vert $\textit{t$\vert
$}=0-1.0~(GeV/c)$^{2}$. In addition, there is a certain scatter of
data of experimental points for close measurements. Of course, it
would be interesting to obtain the analyzing powers of A(d,d')X
reaction on the mentioned targets with other values of the initial
momentum of deuteron and the detection angles of secondary
deuterons.

It has been specified in paper \cite{Ladygin2002} that the deviation of the
$A_{yy}$ value from the predictions of PWIA can be connected with the
contribution of double rescatterings \cite{Kobushkin1998} (thus the rescatterings
and quark exchange considerably influence the polarization observables of
the reaction of inclusive $^{12}$C(d,p) breakup for kinematical region of
high values of momentum in the deuteron), or with the presence of
nonnucleonic degrees of freedom in the deuteron \cite{Azhgirey2000} (when the
influence of baryon resonances as admixture to the deuteron on the momentum
dependences of observables in high-power backward elastic dp- scattering is
considered).

In paper \cite{Ladygin2006} the obtained experimental data of analyzing powers
have been analyzed and it has been stated that data for values $A_{yy}$ at
$\vert $t$\vert  \le $0.8~(GeV/c)$^{2}$ differ from PWIA \cite{Ladygin2002}
calculations for standard DWFs and from the data received in dp- and ed-
elastic scatterings, that is, such behavior and difference indicates to the
sensitivity of $A_{yy}$ to baryonic resonance excitation through
double-collision interactions.

The values of the tensor-tensor and vector-vector polarization transfers can
be calculated as well, respectively \cite{Ladygin2002}:

\begin{equation}
\label{eq19}
K_{yy} = \frac{5T_{11}^2 + T_{00}^2 - 8P^2T_{10}^2 }{T_{00}^2 + 2T_{11}^2 +
4P^2T_{10}^2 };
\end{equation}

\begin{equation}
\label{eq20} K_y = 2\frac{T_{00} T_{11} + 2P^2T_{10}^2 }{T_{00}^2
+ 2T_{11}^2 + 4P^2T_{10}^2 }.
\end{equation}

Unfortunately, we failed to compare these theoretical estimates of
$K_{yy}$ and $K_{y}$ at $a$=0.4 for Nijmegen group potentials
(Figures 5 and 6) with the experimental data, since the latter
were not found in the scientific literature. Therefore, the
experiments on obtaining these polarization observables remain
relevant.

Calculations of the analyzing powers and the polarization transfers for
Nijmegen group potentials (NijmI, NijmII and Nijm93) are compared with the
theoretical estimates for the other three models - DDM \cite{Platonova20101},
fss2 \cite{Fujiwara2001} and OBEPC \cite{Machleidt1989}.

In paper \cite{Rekalo1996} have been calculated the vector
(tensor) transfer polarization coefficients $k_a^{a'} $
($k_{aa}^{a'a'} )$ (with $a=x$, $y$ or $z)$ from initial to final
deuterons for $d + p \to d + X$ process in $\sigma $- and $\omega
$- exchange models. The largest sensitivity of values $k_y^{y'} $
to \textit{$\rho $ }is in the region $p>3$ fm$^{-1}$ and the
position of zero is strongly \textit{$\rho $} dependent.

\begin{center}
\textbf{4. Conclusions}
\end{center}

The polarization observables in the reactions of A(d,d')X type
have been calculated based on the previously obtained coefficients
of the analytical form of the deuteron wave function (\ref{eq3})
in coordinate representation for phenomenological realistic
nucleon-nucleon Nijmegen group potentials (NijmI, NijmII and
Nijm93). Within the framework of the plane-wave impulse
approximation model \cite{Ladygin2002} the theoretical values of
the tensor $A_{yy}$ and vector $A_{y}$ analyzing powers have been
calculated. They are compared with the experimental data of the
reaction of deuterons inelastic scattering on hydrogen, carbon and
beryllium. What is more, the theoretical values of the
tensor-tensor $K_{yy}$ and vector-vector $K_{y}$ polarization
transfers have been estimated, which are sensitive to the
amplitude of NN$ \to $NN*(2190) process up to $\vert
$\textit{t$\vert $}=1.5~(GeV/c)$^{2}$.

In practice, it is sometimes more convenient to use the partial
cross-sections $\sigma _i $ or the spin-flip cross-sections
$\sigma _i^{(sf)} $ \cite{Ladygin2002, Suzuki1994}, which are
characterized by a spin flip and are determined by the very
polarization observables $A_{yy}$, $K_{yy}$, $K_{y}$ obtained in
this paper:

\begin{equation}
\label{eq21}
\left\{ {\begin{array}{l}
 \sigma _0 = \frac{1}{6}\left( {2 + 3K_y + K_{yy} } \right); \\
 \sigma _1 = \frac{1}{9}\left( {4 - (A_{yy} + P_{yy} ) - 2K_{yy} } \right);
\\
 \sigma _2 = \frac{1}{18}\left( {4 + 2(A_{yy} + P_{yy} ) - 9K_y + K_{yy} }
\right); \\
 \end{array}} \right.
\end{equation}

\begin{equation}
\label{eq22}
\left\{ {\begin{array}{l}
 \sigma _0^{(sf)} = \frac{1}{3}\left( { - 1 - 2A_{yy} + 6K_y } \right); \\
 \sigma _1^{(sf)} = \frac{2}{3}\left( {2 + A_{yy} - 3K_y } \right); \\
 \sigma _2^{(sf)} = 0. \\
 \end{array}} \right.
\end{equation}

Finally these polarization observables (\ref{eq21}) can be applied to determination
of the spin structure of $\vec {1} + A \to \vec {1} + B$ process and cross
section \cite{Ohlsen1972}

\begin{equation}
\label{eq23}
I(\theta ,\varphi ) = I_0 (\theta )\left( {1 + \frac{3}{2}\sum\limits_j {p_j
A_j (\theta )} + \frac{1}{3}\sum\limits_{j,k} {p_{jk} A_{jk} (\theta )} }
\right),
\end{equation}

where $p_{l'} $ - the outgoing polarization components:

\begin{equation}
\label{eq24}
p_{l'} I(\theta ,\varphi ) = I_0 (\theta )\left( {P_{l'} (\theta ) +
\frac{3}{2}\sum\limits_j {p_j K_j^{l'} (\theta )} +
\frac{1}{3}\sum\limits_{j,k} {p_{jk} K_{jk}^{l'} (\theta )} } \right);
\end{equation}

$A_j (\theta )$, $A_{jk} (\theta )$ - the analyzing powers;
$P_{l'} (\theta )$ - outgoing polarization (unpolarized incident
beam); $K_j^{l'} (\theta )$, $K_{jk}^{l'} (\theta )$ -
polarization transfer coefficients (tensor-tensor $K_{yy}$ and
vector-vector $K_{y}$ polarization transfers).

The results of numerical calculations of the partial
cross-sections $\sigma _i $ and the spin-flip cross-sections
$\sigma _i^{(sf)} $ at $a$=0.4 are shown in Figure 7.

DWFs for Nijmegen group potentials can be applied for calculations of tensor
and vector analyzing powers and compared with their experimental data for
the deuteron fragmentation reaction with a momentum of 9 GeV/c on the nuclei
of hydrogen and carbon given high proton transverse momentums
\cite{Azhgirey2008}. Calculations need to be conducted within light-front
dynamics with the use of different DWFs (including relativistic DWF).

Moreover, it is of interest to conduct the comparative analysis
between high-precision experimental results \cite{Kurilkin2015}
for vector $A_{y}$ and tensor $A_{yy}$, $A_{xx}$, $A_{xz}$
analyzing powers of \textit{dd}$ \to ^{3}$H$p$ reaction with
deuterons energy of 200 MeV in a full angular range in the
center-of-mass system and theoretical calculations, made within
the multiple-scattering model with the use of standard waves
functions for the three-nucleon bound state and the deuteron.

% graphs

\pdfximage width 100mm {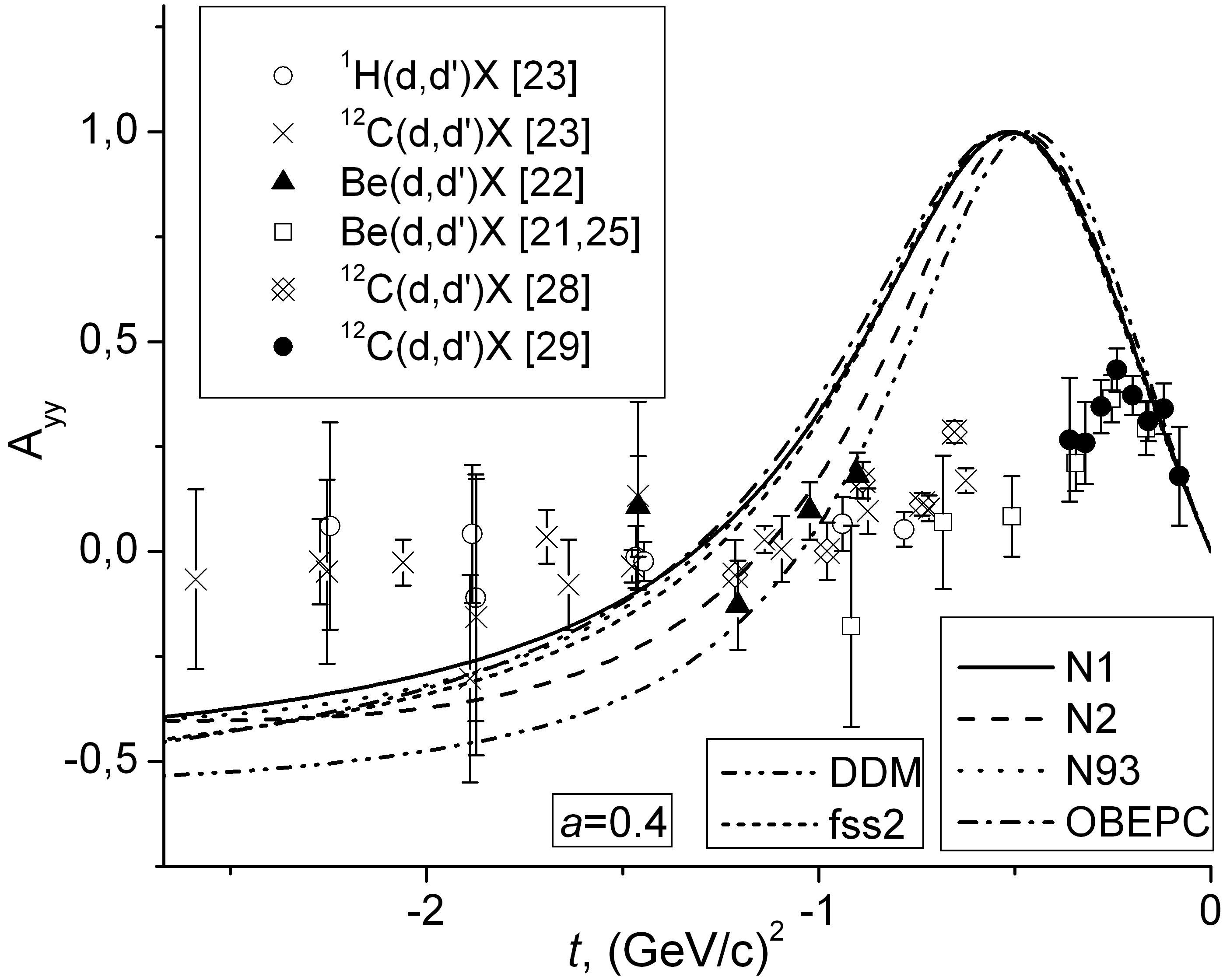}\pdfrefximage\pdflastximage

\textbf{Fig. 1.} Tensor analyzing power $A_{yy}$ at $a$=0.4

\pdfximage width 100mm {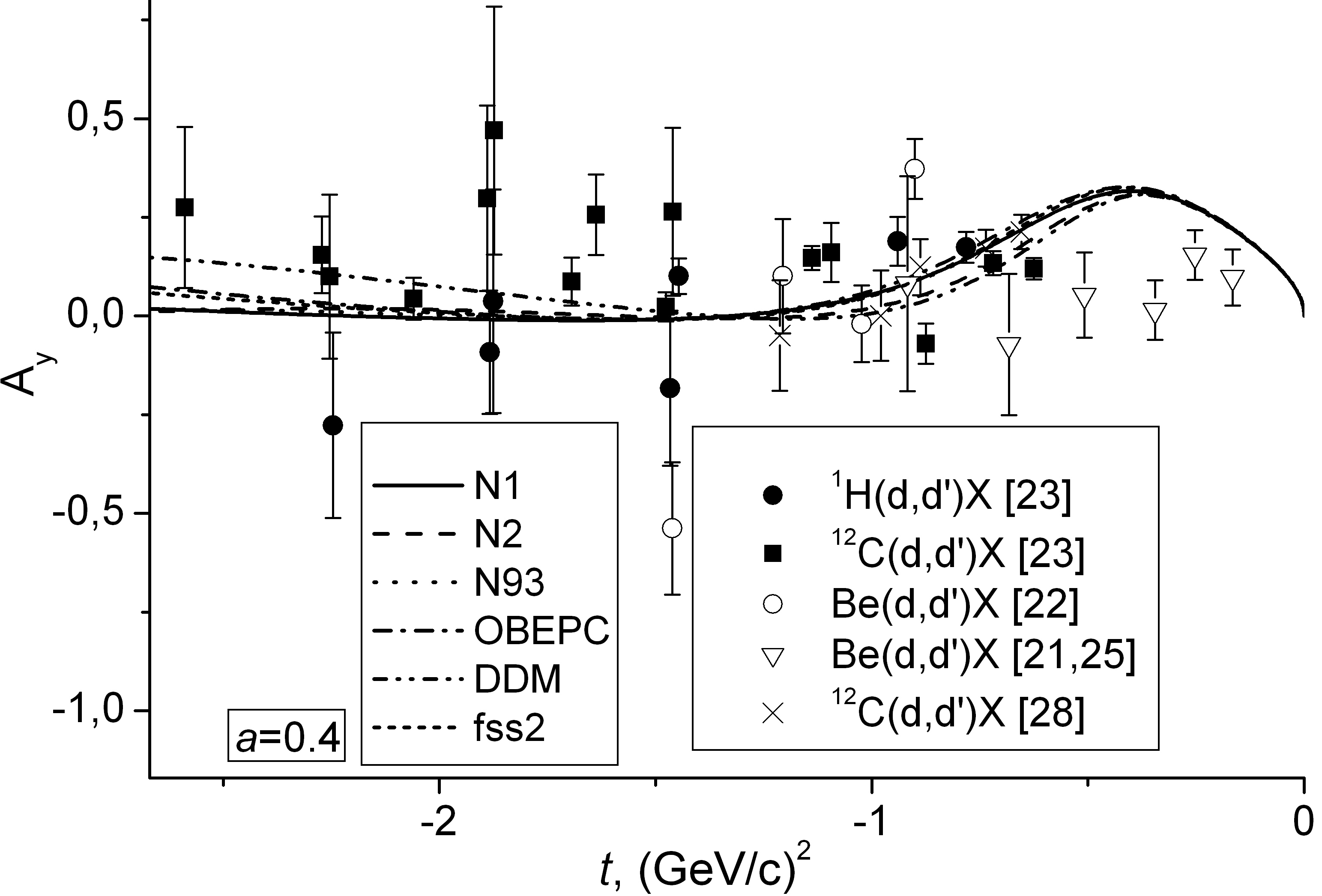}\pdfrefximage\pdflastximage

\textbf{Fig. 2.} Vector analyzing power $A_{y}$ at $a$=0.4

\pdfximage width 100mm {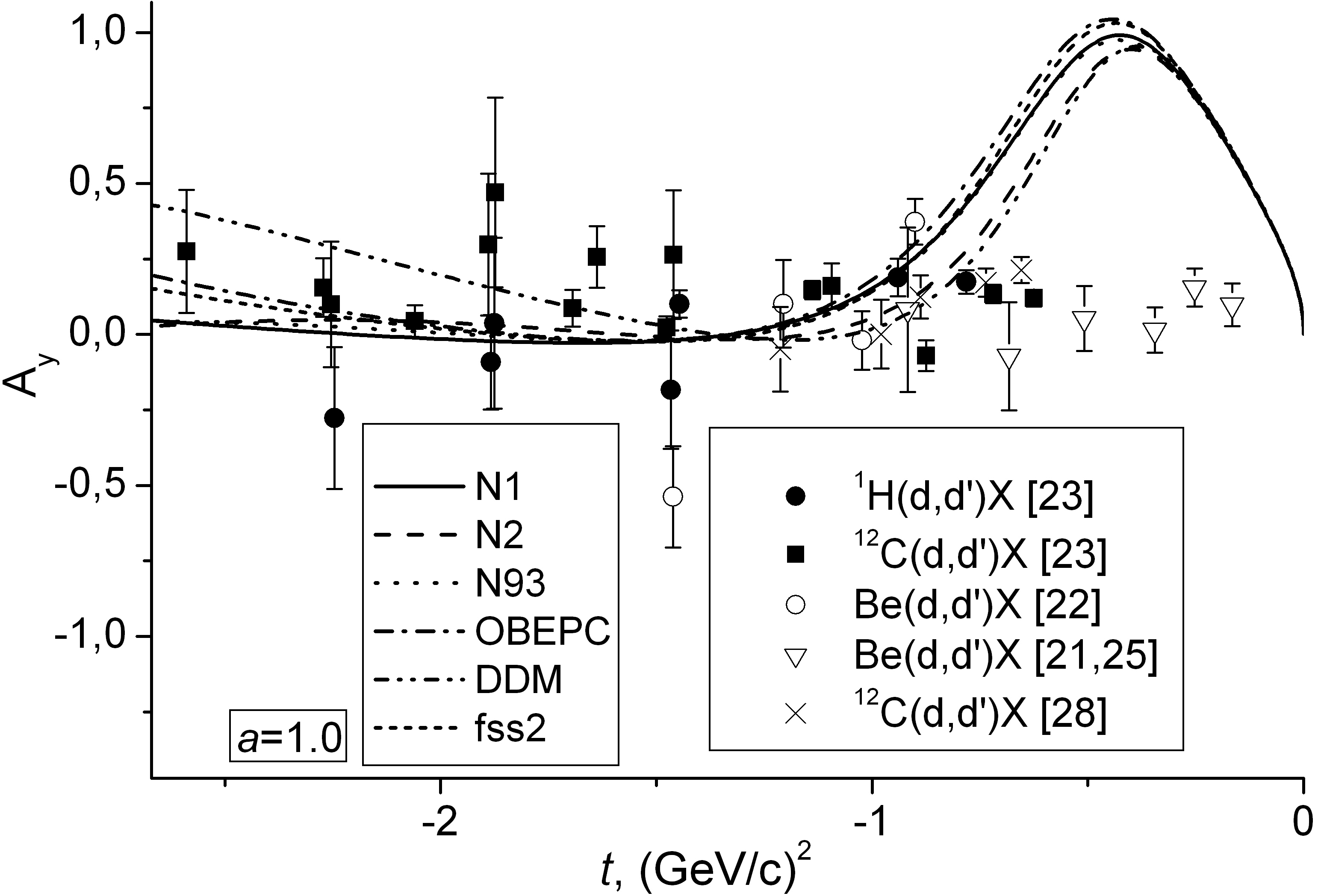}\pdfrefximage\pdflastximage

\textbf{Fig. 3.} Vector analyzing power $A_{y}$ at $a$=1.0

\pdfximage width 100mm {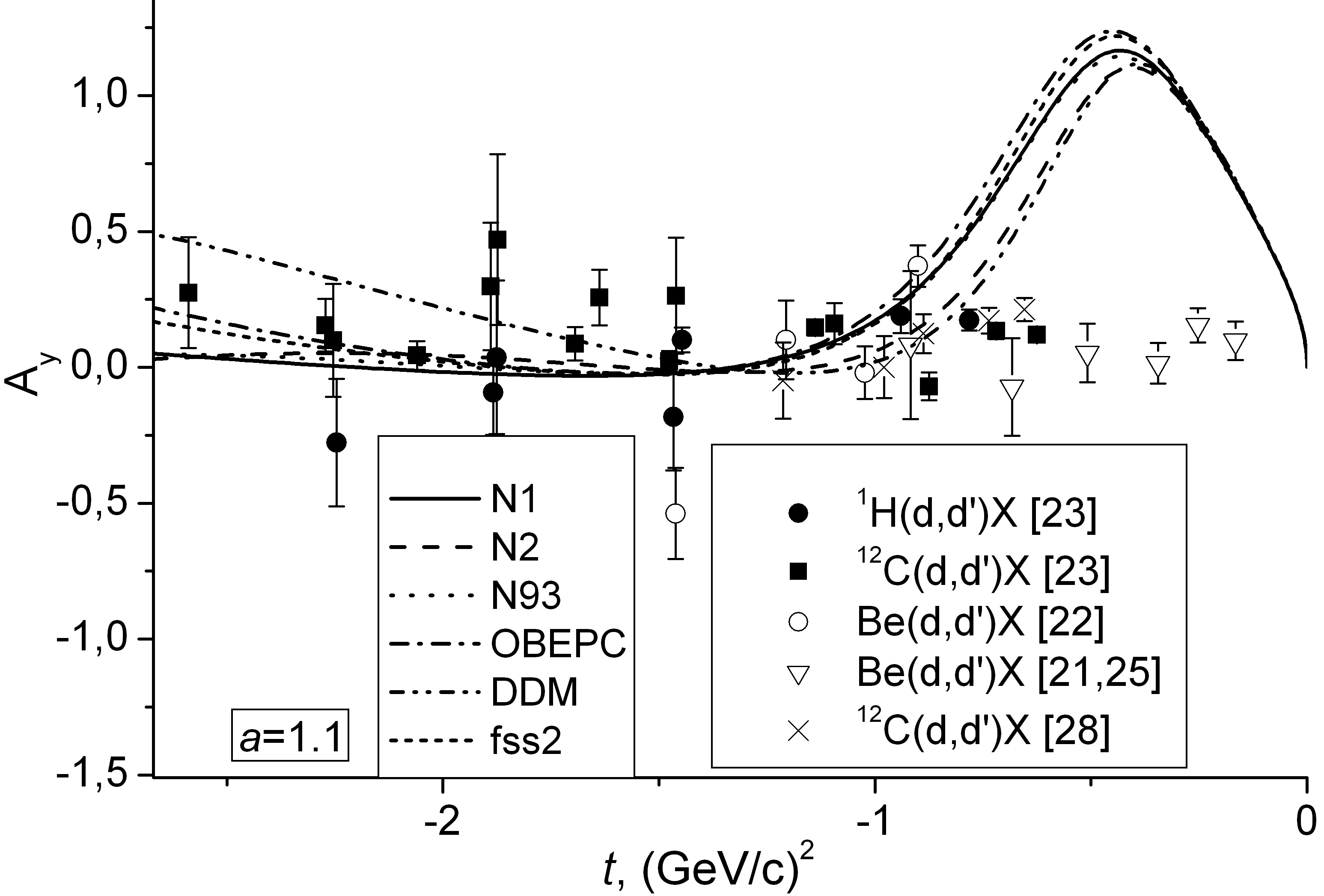}\pdfrefximage\pdflastximage

\textbf{Fig. 4.} Vector analyzing power $A_{y}$ at $a$=1.1

\pdfximage width 100mm {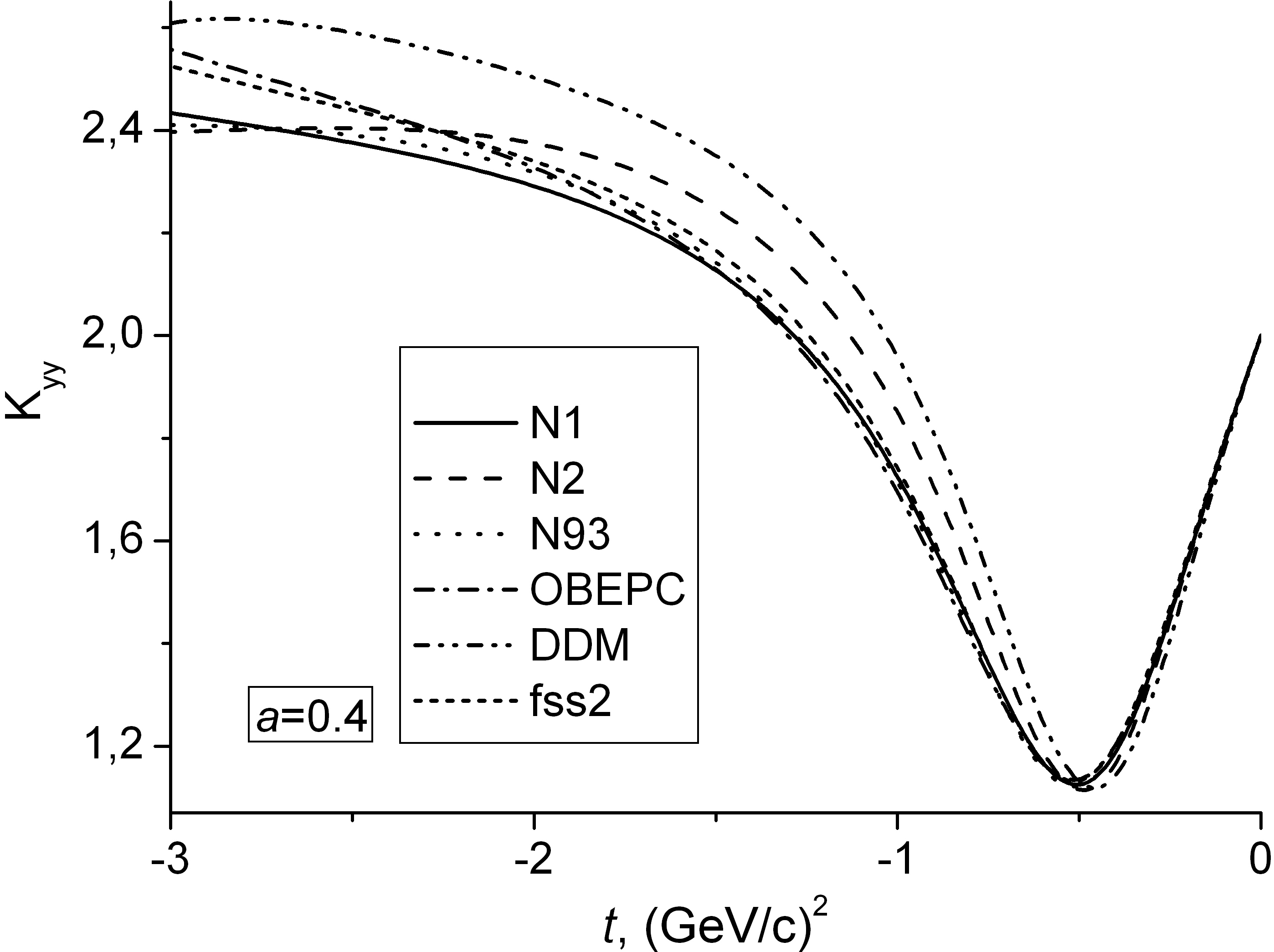}\pdfrefximage\pdflastximage

\textbf{Fig. 5.} Tensor-tensor polarization transfer $K_{yy}$ at
$a$=0.4

\pdfximage width 100mm {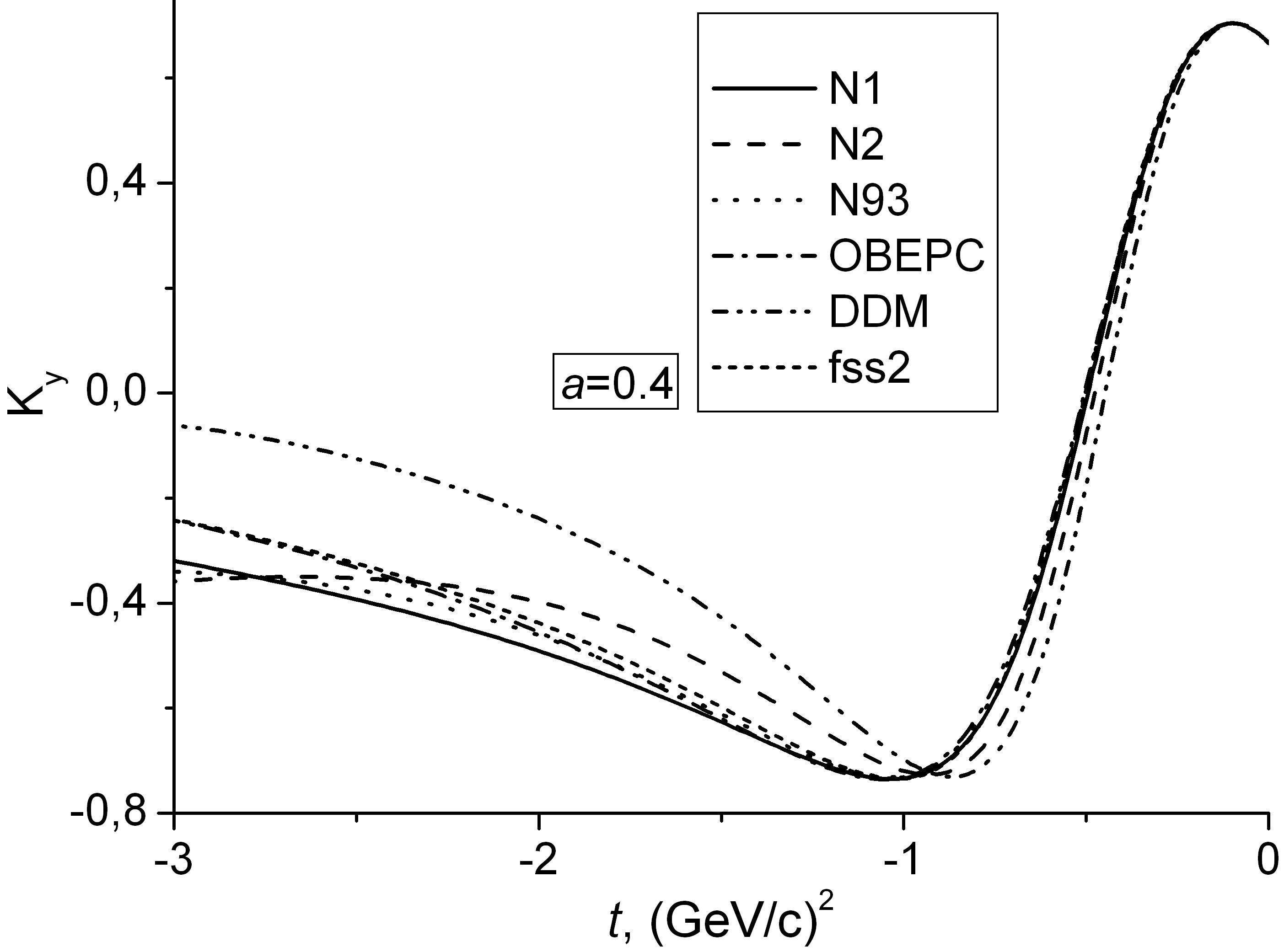}\pdfrefximage\pdflastximage

\textbf{Fig. 6.} Vector-vector polarization transfer $K_{y}$ at
$a$=0.4

\pdfximage width 100mm {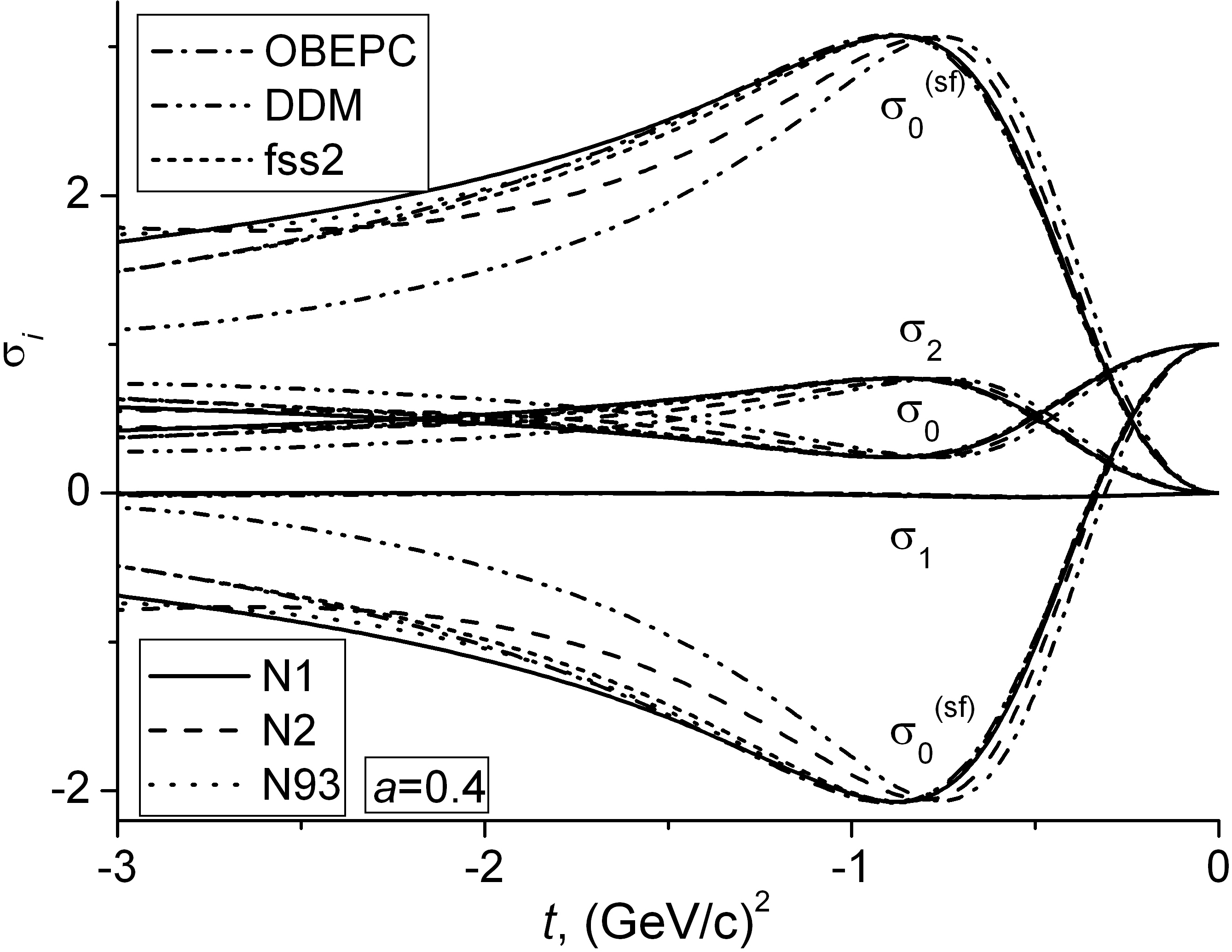}\pdfrefximage\pdflastximage

\textbf{Fig. 7.} Partial cross-sections $\sigma _i $ and spin-flip
cross-sections $\sigma _i^{(sf)} $ at $a$=0.4

\end{document}